\documentclass[12pt]{article}
\usepackage{amssymb}
\usepackage{amssymb}
\usepackage{amsthm}
\usepackage{bm}
\usepackage{amsmath}
\usepackage{amsfonts}
\usepackage{graphicx}
\usepackage{multirow}
\newtheorem{theorem}{Theorem}   
\usepackage{xcolor}
\usepackage{fancyhdr}
\usepackage{hyperref}
\usepackage{color}
\usepackage[margin=1in]{geometry}
\usepackage[authoryear, round]{natbib}

\hypersetup{
    colorlinks =    true,
    citecolor =     blue,
    linkcolor =     blue,
    pdftitle=       {Title},
    pdfsubject=     {Title},
    pdfauthor=      {A. Uthor}
}
      \usepackage[ulem=normalem]{changes}
    
\begin{document}
	
\begin{titlepage}
\title{Forecast Combination Under Heavy-Tailed Errors}
\author{Gang Cheng\thanks{Corresponding author}, \
Sicong Wang, Yuhong Yang}
       
\maketitle

\abstract{Forecast combination has been proven to be a very important technique to obtain accurate predictions. In many applications, forecast errors exhibit heavy tail behaviors for various reasons. Unfortunately, to our knowledge, little has been done to deal with forecast combination for such situations. The familiar forecast combination methods such as simple average, least squares regression, or those based on variance-covariance of the forecasts, may perform very poorly. In this paper, we propose two nonparametric forecast combination methods to address the problem. One is specially proposed for the situations that the forecast errors are strongly believed to have heavy tails that can be modeled by a scaled Student's $t$-distribution; the other is designed for relatively more general situations when there is a lack of strong or consistent evidence on the tail behaviors of the forecast errors due to shortage of data and/or evolving data generating process. Adaptive risk bounds of both methods are developed. Simulations and a real example show superior performance of the new methods.}

\vspace{10pt}


\end{titlepage}


\section{Introduction}

When multiple forecasts are available for a target variable, well designed forecast combination methods can often outperform the best individual forecaster, as demonstrated in the literature of applications of forecast combinations in fields such as tourism, wind power generation, finance and economics in the last fifty years.

Many combination methods have been proposed from different perspectives since the seminal work of forecast combination by \cite{BG1969}. See the discussions and summaries in \cite{C1989}, \cite{NH2002} and \cite{T2006} for key developments and many references. More recently,  
 \cite{L2013}  provided theoretical and numerical comparisons between adaptive and simple forecast combination methods. However, to our knowledge, few studies have proposed/discussed forecast combination methods that target at cases where the forecast errors exhibit heavy tail behaviors. {In this paper, heavy tailed distributions may sometimes loosely refer to distributions with tails heavier than Gaussian distributions, although specific choices such as $t$-distributions will be studied}. In {many} such situations, the familiar forecast combination methods such as simple average, least squares regression with or without constraints, or those based on variance-covariance of the forecasts, may perform very poorly (some numerical examples are provided in sections 4 and 5 in this paper). As a matter of fact, many important variables in finance, economics and other areas {are believed} to have heavy tails. For example, \citet{M2001} discussed the evidences of heavy tailed distributions to model the exchange rates,  and \citet{H2013} modeled the U.S. GDP with a Student's $t$ distribution with a low degrees of freedom. {Therefore, it is practically very useful to design forecast combination methods to handle heavy tailed situations.}

In this paper, we propose two forecast combination methods following the spirit of the AFTER strategy by \cite{Y2004}. One is specially designed for situations when there is strong evidence that the forecast errors are heavy-tailed and can be modeled by a scaled Student's $t$-distribution. The other one is designed for more general uses. For the former case, we assume that the forecast errors follow a scaled Student's $t$-distribution with possibly unknown scaled parameter and degrees of freedom. For situations when the identification of the heaviness of tails of the forecast errors is not feasible, normal, double-exponential and scaled Student's $t$-distributions are considered at the same time as candidates for the distribution form of the forecast errors. In either case, no parametric assumptions are needed on the relationships of the candidate forecasts.  

Technically, if the forecast errors are assumed to follow a normal or a double-exponential distribution with zero mean, then the conditional probability density functions used in the combining process of the AFTER scheme can be estimated relatively easily for all the candidate forecasters because the estimation of the conditional scale parameters is straightforward. See, e.g., \cite{ZY2004} and \cite{WY2012}, for more details. However, this is not thue if a scaled Student's $t$-distribution is assumed. Among the literature discussing the maximum likelihood parameter estimation in Student's $t$-regressions in the last few decades, \cite{FS1999} and \cite{F2008} provided comprehensive summaries of the convergence properties of the parameter estimations in different situations. Both of them showed that the estimation of the degrees of freedom and the scale parameter simultaneously in a scaled Student's $t$-regression models suffers from monotonic likelihood because the likelihood goes to infinity as the scale parameter goes to zero if the degrees of freedom $\nu$ is not large enough. To deal with this difficulty, methods other than maximum likelihood estimation have been proposed in the literature. For example, one may fix the degrees of freedom first then estimate the scale parameter using method of moments or other tools \cite[see, e.g.,][]{KZ2003}. 

In this paper, we follow a two-step procedure to estimate the density function given a forecast error sequence. First, estimate the scale parameter for each element in a given candidate pool of degrees of freedom. Note that each combination of the degrees of freedom and the scale parameter leads to a different estimate of the density function. Second, the weight of a density estimate is assigned from its relative historical performance. The final density estimate is a weighted mean of all the candidate density estimates. More details about this procedure, including how to determine the pool of candidate estimates, are available in section 2. There are three major advantages of this procedure: first,  because a pool of degrees of freedom (rather than a single candidate) is considered,  it reduces the potential risk of picking a degrees of freedom parameter that is far from the truth. Second,  the likelihood that each candidate density estimate is the best is purely decided by data. Third,  the calculation of the combined estimator is easy and fast.  

It is worth pointing out that some popular combination methods in the literature make assumptions on the distributions of forecast errors that do not necessarily exclude heavy tailed  behaviors. For example, methods that are based on the estimation of variance-covariance of forecasters require the existence of variances. Regression based forecast combination methods \cite[see, e.g.,][]{GR1984} assume the existence of certain moments of the forecast errors. However, to our knowledge, these methods are not really designed to handle heavy-tailed errors and are not expected to work well for such situations. 

Prior to our work, efforts have been made to deal with error distributions that have tails heavier than normal by adaptive forecast combination methods. For example, \cite{S2010} assumed that the tails of the target variables are no heavier than exponential decays, which restrict the heaviness of the tails of the forecast errors. \cite{WY2012} designed a method for errors heavier than the normal distributions but not heavier than the double-exponential distributions. However, none of these methods can deal with forecast  errors with tails as heavy as that of Student's $t$-distributions. The new AFTER methods in this paper will be shown to handle such situations.

The plan of the paper is as follows: section 2 introduces the forecast combination method designed for heavy-tailed error distributions; in section 3, a more general combination method is proposed. Simulations are presented in section 4, and section 5 provides a real data example. Section 6 includes a brief concluding discussion. The proofs of the theoretical results are in the appendix.


\section{$t$-AFTER}

In this section, we propose a forecast combination method when there is strong evidence that the random errors in the data-generating process are heavy-tailed and can be modeled by a scaled Student's $t$-distribution.

\subsection{Problem Setting}

Suppose at each time period $i\geq 1$, there are $J$ forecasters available for predicting $y_i$ and the forecast combination starts at $i_0\geq 1$. Note that some combination methods may require $i_0$ to be large enough, e.g., 10, to give reasonably accurate combinations.  Let $\hat{y}_{i,j}$ be the forecast of $y_i$ from the $j$-th forecaster. Let $\hat{Y}_i: = (\hat{y}_{i,1},\cdots,\hat{y}_{i,J})$ be the vector of candidate forecasts for $y_i$ made at time point $i-1$.

Suppose $y_i:=m_i+\epsilon_i$, where $m_i$ is the conditional mean of $y_i$ given all available information prior to observing $y_i$ and $\epsilon_i$ is the random error at time $i$. Assume $\epsilon_i$ is from a distribution with probability density function ($pdf$) $\frac{1}{s_i}h(\frac{x}{s_i})$, where $s_i$ is the scale parameter that depends on the data before observing $y_i$ and $h(\cdot)$ is a $pdf$ with mean $0$ and scale parameter $1$. 

Let $W_i:= (W_{i,1},\cdots,W_{i,J})$ be a vector of combination weights of $\hat{Y}_i$. It is assumed that $\sum_{j=1}^J W_{i,j}=1$ and $W_{i,j}\geq 0$ for any $i\geq i_0$, $1\leq j\leq J$. Let $W_{i_0} = (w_1,\cdots,w_J)$ be the initial weight vector. The combined forecast for $y_i$ from a combination method is:
\begin{equation}\label{eq:1}
   \hat{y}_i = \langle\hat{Y}_i, W_i\rangle,
\end{equation}
where $\langle{\tt a,b}\rangle$ stands for the inner-product of vectors ${\tt a}$ and ${\tt b}$. Specifically, when needed, we use a superscript $\delta$ on each $W_i$ to denote the combination weights that correspond to the method $\delta$. For example, in the following sections, $W_i^{A_2}$ and $W_i^{A_1}$ stand for the
combination weights from the $L_2$- and $L_1$-AFTER methods, respectively. 

\subsection{The Existing AFTER Methods}

As one recent method of adaptive forecast combination, the general scheme of adaptive forecast combination via exponential re-weighting (AFTER) was proposed by \cite{Y2004}. It has been applied and studied in e.g., \cite{Fa2008}, \cite{IK2008}, \cite{S2008}, \cite{AG2010}, and \cite{L2013} and  \cite{Z2013} handled the case that the variable to be predicted is categorical. 

In the general AFTER formulation, the relative cumulative predictive accuracies of the forecasters are used to decide their combining weights. Let $||\mathbf{x}||_1: = \sum_{i=1}^n |x_i|$ be the $l_1$-norm of vector $\mathbf{x}=(x_1,\cdots,x_n)$. 

The general form of $W_i$ for the AFTER approach is:
\begin{equation}\label{eq:2}
W_i = \frac{{\mathbf l_{i-1}}}{||{\mathbf l_{i-1}}||_1}, 
\end{equation}
where ${\mathbf l_{i-1}} = (l_{i-1,1},\cdots,l_{i-1,J})$ and for any $1\leq j\leq J$,
\begin{equation}\label{eq:3}
l_{i-1,j} = w_j\prod_{i'\geq i_0}^{i-1}
\frac{1}{\hat{s}_{i',j}}h\left(\frac{y_{i'}-\hat{y}_{i',j}}{\hat{s}_{i',j}}\right),
\end{equation}
where $\hat{s}_{i',j}$ is an estimate of $s_{i'}$ from the $j$-th forecaster at time point $i'-1$.

Below, the most commonly used AFTER procedures, the $L_2$-AFTER from \cite{ZY2004} and the $L_1$-AFTER from \cite{WY2012}, are {briefly introduced}. 

\noindent \textbf{$L_2$-AFTER} \quad When the random errors in the data generating process follow a normal distribution or a distribution close to a normal distribution, the $L_2$-AFTER is both theoretically and empirically competitive in providing combined forecasts that perform at least as well as any individual forecaster in any \added{performance} evaluation period plus a small penalty. Let $f_N$ be the $pdf$ of $N(0,1)$. To get $W_i^{A_2}$, first use $f_N$ as the $h$ in (\ref{eq:3}), then plug the new ${\mathbf l_{i-1}}$ into (\ref{eq:2}). The $\hat{s}_{i,j}$ used in the $L_2$-AFTER, denoted as $\hat{\sigma}_{i,j}$, is the sample standard deviation of $\{y_{i'}-\hat{y}_{i',j}\}_{i'=1}^{i-1}$ assuming the \added{random} errors are independent and identically distributed.
 
\noindent \textbf{$L_1$-AFTER} \quad Let $f_{DE}$ be the $pdf$ of a double-exponential distribution with scale parameter 1 and location parameter 0. To get $W_i^{A_1}$, one can follow the same procedure for $W_i^{A_2}$ but use $f_{DE}$ as the $h$ in (\ref{eq:3}).  The $\hat{s}_{i,j}$ used in the $L_1$-AFTER, denoted as $\hat{d}_{i,j}$,  is the mean of $\{|y_{i'}-\hat{y}_{i',j}|\}_{i'=1}^{i-1}$. The $L_1$-AFTER method was designed for robust combination when the random errors have occasional outliers. See \cite{WY2012} for details.

\subsection{The $t$-AFTER Methods}

Since the estimation of the degrees of freedom and the scale parameter simultaneously in a scaled Student's $t$-regression setting suffers from certain theoretical difficulties as mentioned in the introduction, we use a different strategy in this paper. Specifically, we take an estimation procedure that has two steps: 

\begin{enumerate}
\item We decide a pool of candidate degrees of freedom with size $K$.  The elements in the pool are considered to be close to the degrees of freedom of the Students' $t$-distribution that describes the random errors well. For each element in the set, we assume it is the true degrees of freedom to estimate the related scale parameter. So we have $K$ sets of estimate for the degrees of freedom and scale parameter pair.
\item For each of the $K$ sets of estimate, we find its probability to be the true one based on the relative historical performances.
\end{enumerate}

This two-step procedure is used in the $t$-AFTER method for forecast combination when the random errors have heavy tails that can be described well by a Students' $t$-distribution.

Let $\Omega:= (\nu_1,\cdots,\nu_K)$ be a set of degrees of freedom for Student's $t$-distributions. The choice of $\Omega$ will be discussed later in this subsection. Let $w_{j,k}$ ($w_{j,k}\geq 0$ and $\sum_{k=1}^K\sum_{j=1}^Jw_{j,k}=1$) be the initial combination weight of the forecaster $j$ under the degrees of freedom $\nu_k$. 

Let the combining weight of $\hat{Y}_i$ from a $t$-AFTER method be $W_i^{A_t}$ and the combined forecast be $\hat{y}_i^{A_t}$. Then, $W_i^{A_t}$ and $\hat{y}_i^{A_t}$ are obtained via the following steps:
\begin{enumerate}
\item Estimate (e.g., by MLE) $s_i$ for each $\nu_k \in \Omega$ and for each candidate forecaster. The estimate for $s_i$ from the $j$-th forecaster given $\nu_k$ is denoted as $\hat{s}_{i,j,k}$.
\item   Calculate $W_i^{A_t}$ and $\hat{y}_i^{A_t}$:
\begin{equation}\label{eq:4}
W_i^{A_t} = \frac{{\mathbf l}_{i-1}^{A_t}}{||{\mathbf l}_{i-1}^{A_t}||_1},  \quad  \hat{y}_i^{A_t} = \langle\hat{Y}_i, W_i^{A_t}\rangle,
\end{equation}
where ${\mathbf l}_{i-1}^{A_t} = (l_{i-1,1}^{A_t},\cdots,l_{i-1,J}^{A_t})$ and for $1\leq j\leq J$ and any $i\geq
i_0+1$,
\begin{equation}\label{eq:5}
l_{i-1,j}^{A_t} = \sum_{k=1}^{K} l_{i-1,j,k}^{A_t} \quad  \text{with \quad $l_{i-1,j,k}^{A_t} = w_{j,k}\prod_{i'\geq
i_0}^{i-1} \frac{1}{\hat{s}_{i',j,k}}f_t\left(\frac{y_{i'} - \hat{y}_{i',j}}{\hat{s}_{i',j,k}} \biggl|\nu_k\right)$},
\end{equation}
where $f_t(\cdot|\nu)$ is the $pdf$ of a Student's $t$-distribution with degrees of freedom $\nu$.
\end{enumerate}

It is assumed that the elements in $\Omega$ are natural numbers for the sake of convenience. In general, when no specific information is available to estimate the size of candidate degrees of freedom efficiently, one can start with a large but relatively sparse pool (say, $\{1,3,5,8,12,15,20,30\}$) and then may narrow it down based on the performances on some training data sets. When there is strong evidence that the tails of the forecast errors are heavy, the size of $\Omega$ can be relatively small, say no more than $3$ or $5$. In this situation, from our experiences, $\Omega = \{1,3\}$ or $\{1,3,5\}$ works well. 

Obviously, when the random errors in the true model follow a scaled Student's $t$-distribution with a known degrees of freedom $\nu$, then $\Omega:= \{\nu\}$. Then (\ref{eq:5}) can be simplified into:
\begin{equation}\label{eq:6}
l_{i-1,j}^{A_t} = w_{j}\prod_{i'\geq i_0}^{i-1} \frac{1}{\hat{s}_{i',j}}f_t\left(\frac{y_{i'} - \hat{y}_{i',j}}{\hat{s}_{i',j}} \biggl|\nu\right),
\end{equation}
where $w_{j}$ is the initial weight of the $j$-th forecaster and $\hat{s}_{i,j}$ is an estimate of $s_i$ from the $j$-th forecaster using all information at and before time point $i-1$ when the true $\nu$ is known.

\subsection{Risk Bounds of the $t$-AFTER}

To avoid potential redundancy, we first give a risk bound on the $t$-AFTER assuming $\nu$ is known. A more general theorem that treats $\nu$ (and even the form of error distribution) as unknown will be given in section 3.

\subsubsection{Conditions}
\noindent{\bf Condition 1}. There exists a constant $\tau>0$ such that for any $i\geq i_0$,
$$
\Pr(\sup_{1\leq j\leq J}|\hat{y}_{i,j}-m_i|/s_i\leq \sqrt{\tau})=1.
$$

\noindent {\bf Condition 2}. These exists a constant $\xi_1>0$ such that for any $i\geq i_0$ and $1\leq j\leq J$:
$$
\Pr\biggl(\frac{\hat{s}_{i,j}}{s_i}\geq \xi_1\biggl) = 1.
$$

\noindent {\bf Condition 2$'$}. These exists a constant $0<\xi'_1<1$ such that for any $i\geq i_0$ and $1\leq j\leq J$:
$$
\Pr\biggl(\xi'_1\leq \frac{\hat{s}_{i,j}}{s_i}\leq \frac{1}{\xi'_1}\biggl) = 1.
$$

Condition 1 holds when the forecast errors are bounded, which is true in many real applications, although it excludes some time series models such as AR(1). It is required for the development of the theorems in this paper. As you can see that this condition does not require $y_i$ to be bounded so that it allows large outliers to occur in the random errors. When the conditional mean of $y_i$ is known to stay in certain range and the related forecasts are relatively restricted, the condition holds. See section 3.1 of \citet{WY2012} for more discussions on this condition.

Condition 2 generally requires that the estimates of the scale parameters are not too small compared to the truth. Condition 2$'$ requires that the estimates of the scale parameters are not too far from the truth in both directions. 

\subsubsection{Risk Bounds for the $t$-AFTER with a Known $\nu$}
Assume the true forecast errors follow a scaled Student's $t$-distribution with a known degrees of freedom $\nu$. Let $\sigma_i$ and $s_i$ be the conditional standard deviation and scale parameter, respectively, of $\epsilon_i$ at time point $i$ and let $\hat{s}_{i,j}$ be an estimator of $s_{i}$ from the $j$-th forecaster. 

Let $q_i = \frac{1}{s_i}f_{t}\left(\frac{y_i-m_i}{s_i}\bigl|\nu\right)$ be the actual conditional error density function at time point $i$ and $\hat{q}_i^{A_t} = \sum_{j=1}^J
W_{i,j}^{A_t}\frac{1}{\hat{s}_{i,j}}f_{t}\left(\frac{\hat{y}_{i,j}-y_i}{\hat{s}_{i,j}}\bigl|\nu\right)$, where $W_{i}^{A_t}$ is defined in (\ref{eq:4}). So, $\hat{q}_i^{A_t}$ is the mixture estimator of $q_i$ from the $t$-AFTER procedure. Let $D(f||g):= \displaystyle\int f\log\frac{f}{g}$ be the Kullback-Leibler divergence between two density functions $f$ and $g$. So, $E\bigl(D(q_i||\hat{q}_i^{A_t})\bigl)$ is a measure of the performances of $\hat{q}_i^{A_t}$ as an estimate of $q_i$ under the Kullback-Leibler divergence at time point $i$.

\begin{theorem}
If the random errors are from a scaled Student's $t$-distribution with degrees of freedom $\nu$ and Condition 2 holds, then:
$$
\frac{1}{n}\sum_{i=i_0+1}^{i_0+n} ED(q_i||\hat{q}_i^{A_t})
\leq 
\inf_{1\leq j\leq J}\left(\frac{\log\frac{1}{w_j}}{n} +
\frac{1}{n}\sum_{i=i_0+1}^{i_0+n}E\frac{(m_i-\hat{y}_{i,j})^2}{2s_i^2} 
+ \frac{B_1}{n}\sum_{i=i_0+1}^{i_0+n}E\frac{(\hat{s}_{i,j}-s_i)^2}{s_i^2}\right).
$$
Further, if $\nu$ is strictly larger than 2 and Conditions 1 and 2$'$ hold, then 
$$
\frac{1}{n}\sum_{i=i_0+1}^{i_0+n} E\frac{(m_i-\hat{y}_i^{A_t})^2}{\sigma_i^2}
\leq 
C\inf_{1\leq j\leq J}\left(\frac{\log\frac{1}{w_j}}{n} + \frac{B_2}{n}\sum_{i=i_0+1}^{i_0+n}
E\frac{(m_i-\hat{y}_{i,j})^2}{\sigma^2_i} 
+ \frac{B_3}{n}\sum_{i=i_0+1}^{i_0+n}E\frac{(\hat{s}_{i,j}-s_i)^2}{s_i^2}\right).
$$
In the above, $C$, $B_1, B_2$ and $B_3$ are constants. $B_1$ and $B_3$ depend on $\xi_1$ and $\xi'_1$, respectively. $B_2$ is a function of $\nu$ and $C$ depends on $\tau$ and $\xi'_1$.
\end{theorem}

\noindent{\bf Remarks.} 
\begin{enumerate}
\item When only Condition 2 is satisfied, Theorem 1 shows that the cumulative distance between the true densities and their estimators from the $t$-AFTER is upper bounded by the cumulative (standardized) forecast errors of the best candidate forecaster plus a penalty that has two parts: squared relative estimation errors of the scale parameters and logarithm of the initial weights. This risk bound is obtained without assuming the existence of variances of the random errors and $\hat{s}_{i,j}/s_i$ is only required to be lower-bounded.
\item When $\nu$ is assumed to be strictly larger than 2 and both Conditions 1 and 2$'$ are satisfied, Theorem 1 shows that the cumulative forecast errors have the same convergence rate of the cumulative forecast errors of the best candidate forecaster plus a penalty that depends on the initial weights and efficiency of scale parameters estimation. The risk bounds hold even if the the distribution of random errors have tails as heavy as $t_3$.
\item If there is no prior information to decide the $w_j$'s in (\ref{eq:6}), then equal initial weights could be applied. That is, $w_j=1/J$ for all $j$. In this case, it is easy to see that the number of candidate forecasters plays a role in the penalty. When the candidate pool is large, some preliminary analysis should be done to eliminate the significantly less competitive ones before applying the $t$-AFTER.
\end{enumerate}

\section{$g$-AFTER}

In section 2, the theoretical risk bounds of the combined forecasts from the $t$-AFTER are provided when the random errors are known to have Student's $t$-distributions. However, the error distribution is typically unknown.

In this section, we propose a forecast combination method, $g$-AFTER, for situations when there is a lack of strong or consistent evidence on the tail behaviors of the forecast errors due to shortage of data and/or evolving data-generating process. A theorem that allows the random errors to be from one of the three popular distribution families (normal, double-exponential, and scaled Student's $t$) is provided to characterize the performance of the $g$-AFTER.

\subsection{The $g$-AFTER Method}
Let the combining weight of $\hat{Y}_i$ from the $g$-AFTER be $W_i^{A_g}$. For any $i>i_0$, $W_i^{A_g}$ and the associated combined forecast $\hat{y}_i^{A_g}$ are:
\begin{equation}\label{eq:7}
W_i^{A_g} = \frac{{\mathbf l}_{i-1}^{A_g}}{||{\mathbf l}_{i-1}^{A_g}||_1}, \quad \hat{y}_i^{A_g} =\langle\hat{Y}_i, W_i^{A_g}\rangle,
\end{equation}
where ${\mathbf l}_{i-1}^{A_g} = (l_{i-1,1}^{A_g},\cdots,l_{i-1,J}^{A_g})$ and for $1\leq j\leq J$,
\begin{equation}\label{eq:8}
l_{i-1,j}^{A_g} = l_{i-1,j}^{A_2} +c_1 l_{i-1,j}^{A_1} +c_2 l_{i-1,j}^{A_t},
\end{equation}
where $l_{i-1,j}^{A_2}$, $l_{i-1,j}^{A_1}$ and $l_{i-1,j}^{A_t}$ are from the $L_2$-, $L_1$- and $t$-AFTERs, respectively and $c_1$ and $c_2$ are non-negative constants that control the relative importances of the $L_2$-, $L_1$- and $t$-AFTERs in the $g$-AFTER. For instance, $c_1$ and $c_2$ can be small when one has evidence that suggests the random errors are likely to be normally distributed.

\subsection{Conditions}

{\bf Condition 3}. Suppose the random errors have zero mean and are from one of the three families (normal, double exponential, and scaled Student's $t$), and there exists a constant $0<\xi_2\leq1$ such that for any $i\geq i_0$, with probability 1, we have
$$
\xi_2 \leq \frac{\hat{s}_{i}}{s_i} \leq \frac{1}{\xi_2}, 
$$
where $s_i$ the actual conditional scale parameter at time point $i$ and $\hat{s}_i$ refers to any estimate of $s_i$ used in the $g$-AFTER.

This condition requires all the estimates of the scale parameters stay in a reasonable range around the true values. For the $j$-th candidate forecaster, $\hat{s}_i$ is $\hat{\sigma}_{i,j}$ when associated with normal errors, is $\hat{d}_{i,j}$ when associated with the double exponential, and is $\hat{s}_{i,j,k}$ when associated with the scaled Student's $t$ with degrees of freedom $\nu_k$, where $\hat{\sigma}_{i,j}$, $\hat{d}_{i,j}$, $\hat{s}_{i,j,k}$ and $\nu_k$ are defined in section 2.2 and 2.3.

\noindent {\bf Condition 4}. When the random errors in the true model follow a scaled Student's $t$-distribution with degrees of freedom $\nu$, assume there exist positive constants $\underline{\nu}$, $\lambda$ and $\bar{\nu}$ such that,
$$
\underline{\nu}\leq \min_{\nu_k\in \Omega}(\nu_k,\nu)-2\leq \bar{\nu}, \quad \max_{\nu_k\in \Omega}|\nu_k-\nu|\leq \lambda.
$$

\subsection{Risk Bounds for the $g$-AFTER}

Let $w_j^{A_2}$ and  $w_j^{A_1}$ be the initial combination weights of the forecaster $j$ in the $L_2$- and $L_1$-AFTERs respectively and $w_{j,k}^{A_t}$ be the initial combination weight of the $j$-th forecaster under the degrees of freedom $\nu_k$ in the $t$-AFTER. 

Let $\hat{W}_{i,j}^{A_2} = \frac{l_{i-1,j}^{A_2}}{||{\mathbf l}_{i-1}^{A_g}||_1}$, $\hat{W}_{i,j}^{A_1} = \frac{c_1l_{i-1,j}^{A_1}}{||{\mathbf l}_{i-1}^{A_g}||_1}$ and $\hat{W}_{i,j,k}^{A_t} = \frac{c_2l_{i-1,j,k}^{A_t}}{||{\mathbf l}_{i-1}^{A_g}||_1}$, where $l_{i-1,j,k}^{A_t}$ is defined in (\ref{eq:5}) and ${\mathbf l}_{i-1}^{A_g}$ is defined in (\ref{eq:8}). So, $\hat{W}_{i,j}^{A_2}$, $\hat{W}_{i,j}^{A_1}$ and $\hat{W}_{i,j,k}^{A_t}$ are the weights of the density estimates under normal, double-exponential and scaled Student's $t$ with degrees of freedom $\nu_k$ in the $g$-AFTER procedure at time point $i-1$ from the $j$-th forecast, respectively. Let $G = \sum_{j=1}^J (w_j^{A_2} + c_1w_j^{A_1} + c_2\sum_{k}w_{j,k}^{A_t})$, where $c_1$ and $c_2$ are defined in (\ref{eq:8}).

Let $q_i$ be the $pdf$ of $\epsilon_i$ at time point $i$ and  its estimator from a $g$-AFTER procedure be:
$$
\hat{q}_{i}^{A_g} = \sum_{j=1}^J\left(\hat{W}_{i,j}^{A_2} \frac{1}{\hat{\sigma}_{i,j}}f_N\left(\frac{\hat{y}_{i,j}-y_i}{\hat{\sigma}_{i,j}}\right) + \hat{W}_{i,j}^{A_1} \frac{1}{\hat{d}_{i,j}}f_{DE}\left(\frac{\hat{y}_{i,j}-y_i}{\hat{d}_{i,j}}\right)+\sum_{k=1}^{K}\hat{W}_{i,j,k}^{A_t} \frac{1}{\hat{s}_{i,j,k}}f_{t}\left(\frac{\hat{y}_{i,j}-y_i}{\hat{s}_{i,j,k}}\bigl|\nu_k\right)\right).
$$

\begin{theorem}
If Conditions 3 and 4 hold, then for $\hat{y}_i^{A_g}$ from a $g$-AFTER procedure, we have:
$$
\frac{1}{n}\sum_{i=i_0+1}^{i_0+n} ED(q_i||\hat{q}_{i}^{A_g}) \leq \inf_{1\leq j\leq J}\biggl(\frac{B_1}{n}\sum_{i=i_0+1}^{i_0+n} E\biggl(\frac{(m_i-\hat{y}_{i,j})^2}{\sigma^2_i}\biggl) 
+ R\biggl),
$$
where
\begin{equation*}
R = 
\begin{cases}
\frac{\log\left(\frac{G}{w_j^{A_2}}\right)}{n}+\frac{B_2}{n}\sum_{i=i_0+1}^{i_0+n}E\frac{(\hat{\sigma}_{i,j}-\sigma_i)^2}{\sigma^2_i}, 
& \text{under normal errors;} \\
\frac{\log\left(\frac{G}{c_1w_j^{A_1}}\right)}{n}+\frac{B_2}{n}\sum_{i=i_0+1}^{i_0+n}E\frac{(\hat{d}_{i,j}-d_i)^2}{d_i^2}, 
& \text{under double-exponential errors;}\\ 
\inf_{1\leq k\leq K}\left(\frac{\log\left(\frac{G}{c_2w_{j,k}^{A_t}}\right)}{n}+\frac{B_2}{n}\sum_{i=i_0+1}^{i_0+n}E\frac{(\hat{s}_{i,j,k}-s_i)^2}{s_i^2}
+ B_3\bigl|\frac{\nu-\nu_{k}}{\nu}\bigl|\right),  
& \text{under scaled $t$ errors.}\\
\end{cases}
\end{equation*}
If Condition 1 also holds, then 
$$
\frac{1}{n}\sum_{i=i_0+1}^{i_0+n} E\frac{(m_i-\hat{y}_i^{A_g})^2}{\sigma_i^2}\leq
C\inf_{1\leq j\leq J}\left( \frac{B_1}{n}\sum_{i=i_0+1}^{i_0+n}
E\left(\frac{(m_i-\hat{y}_{i,j})^2}{\sigma^2_i}\right) 
+ R\right).
$$
In the above, $C$, $B_1$, $B_2$ and $B_3$ are constants depending on $\tau$, $\xi_2$ and parameters in Condition 4.
\end{theorem}

\noindent{\bf Remarks.} 
\begin{enumerate}
\item Theorem 2 provides a risk bound for more general situations compared to Theorem 1. That is, as long as the the true random errors are from one of the three popular families, similar risk bounds hold.
\item When strong evidence is shown that the errors are highly heavy-tailed, $\Omega$ can be very small with only small degrees of freedom and the $c_2w_{j,k}^{A_t}$ in $G$ can be relatively large (relative to $w_j^{A_2}$ and $c_1w_j^{A_1}$). The more information on the tails of the error distributions is available, the more efficient the allocation of the initial weights can be.
\item Specially, when the true random errors have tails significantly heavier than normal and double-exponential, they could be assumed to be from a scaled Student's $t$-distribution with unknown $\nu$ and a (general) $t$-AFTER procedure is more reasonable. In this case, $l_{i-1,j}^{A_g} = l_{i-1,j}^{A_t}$.

Let $q_i = \frac{1}{s_i}f_t\left(\frac{\hat{y}_{i,j}-y_i}{s_i}\right)$ and $\hat{q}_i^{A_t} = \sum_{j,k} \hat{w}_{i,j,k}^{A_t} \frac{1}{\hat{s}_{i,j,k}}f_{t}\left(\frac{\hat{y}_{i,j}-y_i}{\hat{s}_{i,j,k}}\bigl|\nu_k\right)$ and $\hat{w}_{i,j,k}^{A_t}\geq 0$ for all $j$ and $k$. Without assuming Condition 1 is satisfied, it follows for any $n\geq 1$:
$$
 \frac{1}{n}\sum_{i=i_0+1}^{i_0+n} ED(q_i||\hat{q}_{i}^{A_t}) \leq \inf_{1\leq j\leq J}\left(\frac{\log(1/w_{i,j}^{A_t})}{n} + \frac{B_1}{n}\sum_{i=i_0+1}^{i_0+n}
E\frac{(m_i-\hat{y}_{i,j})^2}{\sigma^2_i} + R^{\ast}\right),
 $$
where $w_{j,k}^{A_t}$ is defined the same as that in section 2.3 and 
$$
R^{\ast} = \inf_{1\leq k\leq K}\left(\frac{B_2}{n}\sum_{i=i_0+1}^{i_0+n}E\frac{(\hat{s}_{i,j,k}-s_i)^2}{s_i^2}
+ B_3\bigl|\frac{\nu-\nu_{k}}{\nu}\bigl|\right).
$$
If Condition 1 is also satisfied, then it follows:
$$
\frac{1}{n}\sum_{i=i_0+1}^{i_0+n} E\frac{(m_i-\hat{y}_i^{A_t})^2}{\sigma_i^2} \leq
C\inf_{1\leq j\leq J}\biggl(\frac{\log(1/w_{i,j}^{A_t})}{n} + \frac{B_1}{n}\sum_{i=i_0+1}^{i_0+n}
E\frac{(m_i-\hat{y}_{i,j})^2}{\sigma^2_i} + R^{\ast}\biggl),
$$
where $C$, $B_1$, $B_2$ and $B_3$ are the same as in Theorem 2.
\end{enumerate}

\section{Simulations}

We consider two simulation scenarios,  with candidate forecasters from linear regression models and autoregressive ($AR$) models. Results from the linear regression models show improvements of the $t$- and $g$-AFTERs over the $L_1$- and $L_2$-AFTERs when the random errors have heavy tails. In the $AR$ settings, the $t$- and $g$-AFTERs are compared to many other popular combination methods in various situations, including cases that the forecast errors are with extremely symmetric/asymmetric heavy tails.  We also compared the performances of the $t$- and $g$-AFTERs to other combination methods on the linear regression models and similar results are found. Only representative results are given here.

In this and the following sections, we have the following settings:
\begin{itemize}
\item Use $\Omega = \{1,3\}$. The $t$-AFTER is proposed mostly to be applied when the error terms exhibit very strong heavy-tailed behaviors. When the degrees of freedom of the Student's $t$-distribution gets larger, the $t$-AFTER becomes similar to the $L_1$- or $L_2$-AFTER. Thus a choice of $\Omega$ with relatively small degrees of freedom in the $g$-AFTER should provide good enough adaption capability. In fact, other options for $\Omega$, such as $\Omega = \{1,3,5,8,15\}$ were considered, and similar results were found. 
\item Since it is usually the case that $g$-AFTER is preferred when the users have no consistent and strong evidences to identify the distribution of the error terms from the three candidate distribution families,  we put equal initial weights to the candidate distributions. So $c_1=1$, $c_2=2$, $w_j^{A_1} = w_j^{A_2} = 1/J$ and $w_{j,k}^{A_t} = \frac{1}{2J}$ are used in the $g$-AFTER.  Note that, for example, if there is clear and consistent evidence that the error distribution is more likely to be from the normal distribution family, then putting relatively large initial weights on the $L_2$-AFTER procedure in a $g$-AFTER can be more appropriate than using equal weights.
\item The $\hat{s}_{i,j,k}$'s are the sample median of the absolute forecast errors before time point $i$ from the forecaster $j$ divided by the theoretical median of the absolute value of a random variable with distribution $t_{\nu_k}$. 
\end{itemize}

\subsection{Linear Regression Models}

\subsubsection{Simulation Settings}
There are $p$ predictors $(X_1,\cdots,X_p)$ available and the true model uses the first $p_0$ predictors with coefficients $\beta = (\beta_1,\cdots,\beta_{p_0})$. That is, $Y = \sum_{i=1}^{p_0} X_i\beta_i + \epsilon$. The $p$ candidate forecasters are generated from the following $p$ models: $Y= \beta_0 + X_1\beta_1+e$,  $Y = \beta_0+\sum_{i=1}^2 X_i\beta_i + e$, $\cdots$, $Y = \beta_0+\sum_{i=1}^p X_i\beta_i + e$.  We take $p = 2p_0-1$ for this scenario. Other settings for $p$ and $p_0$ were also considered and they gave similar results.

The $p$ predictors are generated from a multivariate normal distribution with zero mean and covariance matrix $\Sigma$ with sample size $n = 125$. For the entries in $\Sigma$, the diagonal elements are $1$ and off-diagonal elements are $0.8$. The forecasters are generated after the $90$-th observation, and the combination is generated after the $5$th forecasts. Various distributions for the random errors ($\epsilon$) are considered. Note that, we also tried other structures of $\Sigma$, including the ones with $\Sigma_{i,j} = 0.5^{|i-j|}$ and $\Sigma_{i,j} = I(i=j)$ $\forall1\leq i,j\leq p$. The results are similar.

For each set of $\beta$, we generate 200 sets of $(X_1,\cdots, X_p, Y)$ and on each of the 200 sets, we record the $\frac{1}{20}\sum_{i=106}^{125}(m_i-\hat{y}_i)^2$ (Average Squared Estimation Error (ASEE hereafter)) of each combination method, where $\hat{y}_i$ is the forecast of $y_i$ from this method. Note that, since this is a simulation study, the combined forecasts are compared with the conditional means ($m_i$'s) instead of the observations ($y_i$'s) to better compare the competing methods. For each competing method, the mean ASEE over the 200 data sets is recorded. 

We sample $\beta$ for 200 times independently from a $Unif[1,3]$ for each component with size $p_0$, so 200 sets of mean ASEEs are recorded. In order to compare the performances of the four AFTER based methods, the $L_2$-, $L_1$-, $t$- and $g$-AFTERs, for each $\beta$, the ratios of the mean ASEEs of the $L_2$-, $t$- and $g$-AFTERs over the mean ASEE of the $L_1$-AFTER is recorded. The summaries (means and their standard errors) of the 200 sets of ratios are presented.

\subsubsection{Results}
Three sets of results ($p_0=3, 5 ,10$ respectively) are presented in Table \ref{Table:1} in this subsection. In this table, $A2$, $At$ and $Ag$ stand for the ratios of the mean ASEEs of the $L_2$-, $t$- and $g$-AFTERs over those of the $L_1$-AFTER. The information in the first and second rows indicate the distributions of $\epsilon$: $t_3$ with $\sigma^2=9$ means $\epsilon \sim kt_3$ with $Var(kt_3)=9$. The top numbers in rows 4-6, 8-10 and 12-14 are the mean of the 200 ratios. The numbers in the parentheses are the standard errors of the statistics above them. Rows $3$, $7$ and $11$ tell the number of predictors used in the true models. $DE$ stands for double-exponential with zero mean hereafter.

\subsubsection{Summary}

From Table \ref{Table:1}, in the linear regression setting, we see that the overall performances of the $t$- and $g$-AFTERs are relatively more robust than that of the $L_1$- and $L_2$-AFTERs. Specifically:
\begin{enumerate}
\item When the random errors have heavy tails, the $t$- and $g$-AFTERs provide more accurate forecasts than the $L_2$- and  $L_1$-AFTERs consistently.
\item When the tails of the random errors distributions are not or only mildly heavy, say a normal or a scaled Student's $t$-distribution with a large degrees of freedom, the $g$-AFTER is better than the $t$-AFTER in terms of forecast accuracy.
\item The $L_1$-AFTER outperforms the $L_2$-AFTER when the random errors have heavy tails while $L_2$-AFTER is more accurate than the $L_1$-AFTER when the random errors are not heavy-tailed.
\end{enumerate}

\subsection{AR Models}

\subsubsection{Simulation Settings}
Let the true model be a $AR(p_0)$ process with random errors from certain distributions and the candidate forecasters be based on $AR(1), AR(2),\cdots, AR(p)$ ($1\leq p_0\leq p$), respectively. For results on asymptotically optimal model selection for $AR$ models, see, e.g., \cite{I2007} and \cite{I2012}. We here compare forecast combination methods. 

In this scenario, given $p$, $p_0$ is randomly sampled from a Uniform distribution on $\{1,2,\cdots,p\}$.  Given $p_0$, $\beta$ in the true model is generated; given $\beta$, 200 samples with size $n=125$ from the true model are generated. On each data sample, the candidate forecasters are generated after the $90$-th observation and the ASEE of the last 20 forecasts is recorded. Also, the combined forecasts are compared with the conditional means instead of the observations. For each $\beta$, the mean ASEE of each combining method over the 200 samples is recorded and ratios of the mean ASEEs of other methods over that of the $L_1$-AFTER are recorded. 

We replicate the generation of $p_0$'s (and $\beta$'s) for 200 times and report the mean and its standard error of the 200 ratios for each combination method.

Only the results of $p=5$ are presented (other choices, such as $p=8$ and $10$, provide similar results). 

\subsubsection{Other Combination Methods}
Some other popular combination methods are included in this part and compared with the newly proposed methods. Simple average combination strategy ($SA$) uses the average of the candidate forecasts as the combined forecasts. The $MD$ and $TM$ strategies use the median and the trimmed mean (remove the largest and smallest before averaging) of candidate forecasts, respectively. The variance-covariance estimation based combination method (denoted as $BG$ because it was first proposed by \citet{BG1969}) we use in this paper is the version in \cite{H2008}. Also, a modified $BG$ method with a discount factor $0<\rho<1$ is considered and the results of multiple $\rho$'s are presented. In the modified $BG$, the estimate of the (conditional) variance of the forecast errors of a forecaster at any time point is the associated discounted mean squared forecast error with factor $\rho$.   See, e.g, \citet[][]{SW2006}, for more details.  Hereafter, for example,  $BG_{0.9}$ denotes a $BG$ method with $\rho=0.9$. Two linear-regression based combination methods are also considered: one is the combination via ordinary linear regression ($LR$) and the other one is a constrained  linear regression ($CLR$) combination. The constraints of the $CLR$ are:  all coefficients are non-negative and the sum of the coefficients is 1 (without intercept in the regressions). 

\subsubsection{Results}

Tables \ref{Table:2} and \ref{Table:3} provide the summaries of the simulation results. In these two tables, $A2$, $At$, $Ag$, $SA$, $MD$, $TM$, $BG$, $LR$ and $CLR$ stand for the relative performances of these methods over that of the $L_1$-AFTER. The other entries are defined as in Table \ref{Table:1}. 

Table \ref{Table:2} presents the results for the cases that the random errors are not  (or only mildly) heavy-tailed, while Table \ref{Table:3} contains the results when the random errors have significant heavy tails.

\subsubsection{Summary}

In the autoregression scenario, we see that the $t$- and $g$-AFTERs consistently outperform all other non-AFTER based combination methods in all the simulated situations (heavy tailed or not) and outperform the $L_1$- and $L_2$-AFTERs when the random errors are not normal. Below are some important details:
\begin{enumerate}
\item In between the $t$- and $g$-AFTER,  the latter is more robust since its performances under all scenarios are the best or close to the best. For the $t$-AFTER, its advantages over the $L_1$- and $L_2$-AFTERs are clear when the tails of the distributions of the random errors get heavier.
\item In both Tables \ref{Table:2} and \ref{Table:3}, the $CLR$ is the most competitive method outside the AFTER family. It is because the constraints in the $CLR$ make its weights relatively more stable and resistant to dramatic changes. The $CLR$ gets more competitive when the random errors have heavier tails.
\item The $SA$ and $TM$ are vulnerable to outliers, which hurts their overall performances.  We can see this from both tables. 
\item In our settings, similar to many real application situations, since some of the candidate forecasters are highly correlated, using only the conditional variances to assign relative combining weights may not be enough. This explains why the $BG$ and the discounted $BG$'s are not quite competitive as seen in Tables \ref{Table:2} and \ref{Table:3}.
\end{enumerate}

\section{Real Data Example}

The M3-competition data contain 3003 financial/economical variables in which 1428 (N1402-N2829) have 18 forecasts and the rest have only 6 or 8 forecasts. For each of the 3003 variables, notice that the forecasts are generated all at once (1-, 2-,$\cdots$ and up to 6, 8 or 18-step ahead) by each forecaster. There were 24 candidate forecasters for each of the variables.  We use the 1428 variables with 18 forecasts to conduct the simulation study because some combination methods need a few forecasts to train the parameters before achieving a reasonable level of reliability.

\subsection{Data and Settings}

Let  $\hat{y}_{i'}$ be the forecast of $y_{i'}$ for $n_0\leq i'\leq n_1$, then the mean squared forecast error (MSFE) is $\frac{1}{n_1-n_0+1}\sum_{i=n_0}^{n_1} (y_i - \hat{y}_i)^2$. We use the mean squared forecast errors to measure the prediction performances of the combination methods on each of the 1428 variables. For each variable, the MSFE of each of the other combination methods over the MSFE of the $SA$ is reported. 

Specifically, using the same notations as those in section 4.2,  the averaged relative performances (MSFE) of the $MD$, $TM$, $BG$, discounted $BG$'s, $A2$, $A1$, $At$ and $Ag$ over the $SA$ over the 1428 variables are presented. The main reason that we use the $SA$ as the benchmark on this real data set is that the $SA$ is one of the most popular combination methods with a great reputation in a broad range of applications. Since there are too many candidate forecasters compared to the forecast periods available, the two linear regression related combination methods discussed in section 4.2 are not considered here.  

For each of the variables with 18 forecast periods, the combination starts after the $6$-th forecasts and the MSFE of the last 9 forecasts of each method is recorded for performance comparisons. For each variable, the MSFE ratio of each method over that of the $SA$ is reported. The summaries, mean (and its standard error), median, minimum, the 1$st$, 3$rd$ quartiles (denoted as $Q_1$ and $Q_3$, respectively) and maximum,  of the 1428 ratios of each method are reported in Table \ref{Table:4}.  

Also, the comparison on a subset of M3-competition data is provided. On this subset, the variables are considered to have high potentials to be heavy tailed. For each of the 1428 variables with 18 forecast periods, there are some training data (about 70-128 months). We modeled the training data to find the ones with high potential to have heavy tailed errors. Specifically, let $y_t$ be the observed value of a variable at time $t$ and we fit each variable with a model as: $y_t = \beta_0 + \sum_{j=1}^11\beta_j I(m_t = j) + \beta_{12}y_{t-1} + \cdots + \beta_{16}y_{t-5}$ using AIC in backward selection and the ones with kurtosis larger than 3 are considered to have heavy tails. There are 199 out of 1428 variables are selected.

On the heavy tailed subset, we want to focus on the comparison between the {\tt g-AFTER} and the non-{\tt AFTER} methods because the comparison inside {\tt AFTER} family is well addressed in simulation settings. The reason we choose the {\tt g-AFTER} instead of the {\tt t-AFTER} for further comparison is because {\tt g-AFTER} is practically more efficient since it performs well even the signal of heavy tails is not extremely strong.  So, on this subset, the benchmark method is the {\tt g-AFTER} and the results are reported in \ref{Table:5}.

\subsection{Summary}

\begin{enumerate}
\item From Table \ref{Table:4}, the overall performances of the AFTER based methods are better than the other popular combination methods considered. It also shows that the AFTERs can occasionally be significantly worse than the $SA$ and other methods. 
\item From Table \ref{Table:4},  it is worth noticing that the performances of the AFTERs can be a thousand times better while only about 10 times worse than that of $SA$. An examination reveals that for certain variables, such as N1837 and N2217, some candidate forecasters are consistently and significantly worse than others. In this situation, since the $SA$ can not remove the extreme `disturbing' ones before averaging, its performance is extremely poor. However, the AFTERs essentially ignore the `unreasonable' candidate forecasts so they can be significantly better than the $SA$. 
\item Table \ref{Table:4} suggests that the $t$- and $g$-AFTERs have competitive performances in general while being more robust than others since their overall performances are outstanding and are still acceptable for the worst cases.
\item From the comparison in Table \ref{Table:5}, the $g$-AFTER is significantly better than the non-AFTER methods when the random errors are suspected to have heavy tails. So the robustness of $g$-AFTER is supported by the M3-Competition data.
\end{enumerate}

\section{Conclusions}

Forecast combination is an important tool to achieve better forecasting accuracy when multiple candidate forecasters are available. Although many popular forecast combination methods do not necessarily exclude heavy tailed situations, little is found in the literature that examines the performances of forecast combination methods in such situations with theoretical characterizations.

In this paper, we propose combination methods designed for cases when forecast errors exhibit heavy tail behaviors that can be modeled by a scaled Student's $t$-distribution and for the cases when the heaviness of the forecast errors is not easy to identify. The $t$-AFTER models the heavy-tailed random errors with scaled Student's $t$-distributions with unknown (or known) degrees of freedom and scale parameters. A candidate pool of degrees of freedom are proposed to solve the estimation problem and the resulting $t$-AFTER works well as seen in simulation and real example analysis.

However,  in many cases the heaviness of the tails of the random errors is difficult to identify.  Therefore, we design a combination process for general use and call it $g$-AFTER. For these situations, instead of assuming a certain distribution form for the random errors, a set of possible heaviness of the tails are considered and the combination process automatically decides which ones are more reasonable by giving them high weights. The numerical results suggest the performance of the $g$-AFTER is more robust than other popular combination methods because of its adaptive capability. The design of  the $g$-AFTER provides a general idea: when there are multiple reasonable candidate distributions for the random errors, combining them in an AFTER scheme like the $g$-AFTER for forecast combination should work well. 

\section{Acknowledgement}

This work is partially supported by National Science Foundation grant DMS-1106576.

\section*{Appendix}

\subsection*{A.1}

In this subsection, some simple facts are given. They are used in A.2 of the appendix.
\begin{itemize}
\item Fact 1: $1-(1-t)^a\leq \displaystyle\frac{at}{1-t}$ for $a\geq 0, 0\leq t <1$. Let $f(t,a) = 1-(1-t)^a-at/(1-t)$,
then $f(t,a)\leq 0$ since $\partial f/\partial t = a(1-t)^{-2}((1-t)^{a+1}-1)\leq 0$ and $f(0,a)=0$.
\item Fact 2: $\log(x)\leq x-1$ for $x\geq0$.
\item Fact 3: For any $c>0$, $B(a,b)/B(a,b+c)$ decreases as $b$ increases. The proof is pure arithmetics and the key
point is using the fact that $B(x,y) = \frac{x+y}{x y} \prod_{n=1}^\infty \left( 1+ \dfrac{x y}{n
(x+y+n)}\right)^{-1}$.
\item Fact 4: $E(1+\frac{Y^2}{\nu})^{-1} = \nu/(\nu+1)$, where $Y\sim t_{\nu}$ conditional on $\nu$. Let $Z = Y
\sqrt{(\nu+2)/\nu}$, then it
is easy to show that $E(1+\frac{Y^2}{\nu})^{-1} = B(1/2,(\nu+2)/2)/B(1/2,\nu/2)=\nu/(\nu+1)$.
\item Fact 5: $(s^2-1)/2-\log(s) \leq \frac{s_0+2}{2s_0}(1-s)^2$ if $s\geq s_0> 0$. Using fact 2 to show
that $-\log(s) = \log(1+(1-s)/s)\leq (1-s)/s$.
\end{itemize}

\subsection*{A.2}

{\it Lemma 1} \quad Let $h_{\nu}(x)$ be the density function of $t_{\nu}$, $\underline{\nu}>0$ and
$\lambda>0$ be constants. Then for any $0<s_0\leq s$, $\underline{\nu}\leq \min(\nu,\nu')-2\leq
\bar{\nu}$ and $|\nu-\nu'|\leq \lambda$, we have
\begin{equation*}
\int h_{\nu}(x)\log\frac{h_{\nu}(x)}{\frac{1}{s}h_{\nu'}\bigl(\frac{x-t}{s}\bigl)}\leq C_1(1-s)^2+ C_2t^2 +
C_3\left|\frac{\nu'-\nu}{\nu}\right|,
\end{equation*}
where $C_1$, $C_2$ and $C_3$ are constants depending on $s_0$, $\underline{\nu}$, $\bar{\nu}$ and $\lambda$.\\
Proof: After a proper reorganization, we have
\begin{align*}
E\log\frac{h_{\nu}(X)}{\frac{1}{s}h_{\nu'}\bigl(\frac{X-t}{s}\bigl)} & = 
\log(s) + \frac{1}{2}\log\frac{\nu'}{\nu}
+ \log\frac{B(\frac{1}{2},\frac{\nu'}{2})}{B(\frac{1}{2},\frac{\nu}{2})} \\
& + E\left( 
\frac{1+\nu'}{2}\log\bigl(1+\frac{(X-t)^2}{s^2\nu'} \bigl)
-\frac{1+\nu}{2}\log\frac{X^2+\nu}{\nu}\right)
\end{align*}
\begin{itemize}
\item Let $\nu^{\ast} = \min(\nu, \nu')$ and using the Facts 1, 2 and 3, then:
\begin{align*}
&\quad \log\frac{B(\frac{1}{2},\frac{\nu'}{2})}{B(\frac{1}{2},\frac{\nu}{2})} \leq
\frac{|B(\frac{1}{2},\frac{\nu}{2})-B(\frac{1}{2},\frac{\nu'}{2})|}{B(\frac{1}{2},\frac{\nu}{2})} 
= \frac{\int t^{-1/2}(1-t)^{\nu^{\ast}/2-1}(1-(1-t)^{|\nu-\nu'|/2})dt}{B(\frac{1}{2},\frac{\nu}{2})} \\
& \leq \frac{\frac{|\nu-\nu'|}{2}\int t^{1/2}(1-t)^{\nu^{\ast}/2-2}dt}{B(\frac{1}{2},\frac{\nu}{2})}
= \frac{|\nu-\nu'|}{2} \frac{B(\frac{3}{2},\frac{\nu^{\ast}-2}{2})}{B(\frac{1}{2},\frac{\nu}{2})}
= \frac{|\nu-\nu'|}{2} \frac{B(\frac{3}{2},\frac{\nu^{\ast}-2}{2})}{B(\frac{1}{2},\frac{\nu^{\ast}-2}{2})}
\frac{B(\frac{1}{2},\frac{\nu^{\ast}-2}{2})}{B(\frac{1}{2},\frac{\nu}{2})}\\
&=
\frac{|\nu-\nu'|}{2}\frac{1}{\nu^{\ast}-1}\frac{B(\frac{1}{2},\frac{\underline{\nu}}{2})}{B(\frac{1}{2},\frac{\underline
{\nu}+2}{2})}
=
\frac{|\nu-\nu'|}{\nu}\frac{\nu}{\nu^{\ast}-1}\frac{B(\frac{1}{2},\frac{\underline{\nu}}{2})}{B(\frac{1}{2},\frac{
\underline{\nu}+2}{2})}
\leq \frac{|\nu-\nu'|}{\nu}
\frac{\underline{\nu}+\lambda}{\underline{\nu}+1}\frac{B(\frac{1}{2},\frac{\underline{\nu}}{2})}{B(\frac{1}{2},\frac{
\underline{\nu}+2}{2})}\\
& \leq \frac{|\nu-\nu'|}{\nu}
\frac{\underline{\nu}+\lambda}{\underline{\nu}+1}
\end{align*}
\item Using Fact 2 in A.1, it follows: $\frac{1}{2}\log\frac{\nu'}{\nu}\leq \frac{1}{2}\frac{\nu'-\nu}{\nu}\leq
\frac{1}{2}\frac{|\nu'-\nu|}{\nu}.
$ 
\item It is easy to show that:
\begin{align*}
& \quad E\left\{\log(s) + \frac{1+\nu'}{2}\log\bigl(1+\frac{(X-t)^2}{s^2\nu'} \bigl)
-\frac{1+\nu}{2}\log\bigl(1+\frac{X^2}{\nu}\bigl)\right\} \\
& = E\left\{\log(s)-(1+\nu')\log(s) + \frac{1+\nu'}{2}\log\biggl(\frac{s^2+\frac{(X-t)^2}{
\nu'} }{1+\frac{X^2}{\nu}}\biggl) +\frac{\nu'-\nu}{2}\log\bigl(1+X^2/\nu\bigl)\right\} \\
& \leq -\nu'\log(s) + E\left\{\frac{1+\nu'}{2}\frac{s^2-1+(X-t)^2/\nu' - X^2/\nu}{1+X^2/\nu} +
X^2|\nu'-\nu|/\nu\right\}\\
& \leq (2+\bar{\nu})\frac{2+s_0}{2s_0}(1-s)^2+\frac{\underline{\nu}+3}{\underline{\nu}+2}t^2+C_3^{\ast}
\frac {|\nu'-\nu|} {\nu},
\end{align*}
where $C_3^{\ast}$ is a constant depending on $s_0$, $\underline{\nu}$, $\bar{\nu}$ and $\lambda$.
\end{itemize}
The proof can be completed by combining these steps.

Note that if $\nu$ is known, then $\nu = \nu'$. Then,
$$
E\log\frac{h_{\nu}(X)}{\frac{1}{s}h_{\nu'}\bigl(\frac{X-t}{s}\bigl)} \leq \nu\frac{2+s_0}{2s_0}(1-s)^2+\frac{1}{2}t^2.
$$\\

{\it Lemma 2} \quad Let $h(x)$ be the density function of a double-exponential distribution with $\mu=0$ and
$d=1$, then for $s_0>0$ and $s\geq s_0$ it follows:
\begin{equation*}
\int h(x)\log\frac{h(x)}{\frac{1}{s}h\left(\frac{x-t}{s}\right)}\leq C_4 (1-s)^2+C_5t^2,
\end{equation*}
where $C_4$ and $C_5$ are constants depending only on $s_0$.\\
Proof: since $h(y) = \frac{1}{2}\exp(-|y|)$ and $\exp(-x)\leq 1-x+\frac{x^2}{2}$ for
$x\geq 0$, then
\begin{align*}
E\log\frac{h(Y)}{\frac{1}{s}h\bigl(\frac{Y-t}{s}\bigl)}dy 
& = \log(s) + E\biggl(\frac{|Y-t|}{s}\biggl)-E|Y|
= \log(s) + \frac{\exp(-t) + t}{s}-1\\
& \leq (s-1) + \frac{1+t^2/2}{s}-1 = \frac{t^2}{2s} + (1-s)^2\frac{1}{s}
\leq \frac{t^2}{2s_0} + \frac{1}{s_0}(1-s)^2.
\end{align*}

{\it Lemma 3} \quad \quad Let $h(y)$ be the density function of a standard normal distribution, then for $s_0>0$ and
$s\geq s_0$ it follows:
\begin{equation*}
\int h(x)\log\frac{h(x)}{\frac{1}{s}h\left(\frac{x-t}{s}\right)}\leq C_6 (1-s)^2+C_7t^2,
\end{equation*}
where $C_6$ and $C_7$ are constants depending only on $s_0$.\\
Proof: using Fact 2,
\begin{align*}
E\log\frac{h(Y)}{\frac{1}{s}h\bigl(\frac{Y-t}{s}\bigl)}dy & = \log(s) + \frac{1+t^2-s^2}{2s^2}
= \frac{1}{2s^2}t^2 + \log(s) + \frac{1-s^2}{2s^2} \leq \frac{1}{2s^2}t^2 + (s-1) + \frac{1-s^2}{2s^2}\\
& =\frac{1}{2s^2}t^2 + \frac{2s+1}{2s^2}(s-1)^2 \leq \frac{1}{2s_0^2}t^2 + \frac{2s_0+1}{2s_0^2}(s-1)^2.
\end{align*}

\subsection*{A.3}

In this subsection, we prove Theorem 1. 

Conditional on the information available until time point $i$, it is assumed that $\frac{Y_i-m_i}{s_i}\sim t_{\nu}$,
where $s_i$ is the conditional scale parameter at time $i$. Let $\hat{s}_{i,j}$ be the estimator of $s_i$ from the
$j$-th forecaster.

Let $f^n = \prod_{i=i_0+1}^{i_0+n}\frac{1}{s_i}h\left(\frac{y_i-m_i}{s_i}\right)$ and $q^n = \sum_{j=1}^K
\pi_j\prod_{i=i_0+1}^{i_0+n}\frac{1}{\hat{s}_{i,j}}h\left(\frac{y_i-\hat{y}_{i,j}}{\hat{s}_{i,j}}\right)$,
where $h(\cdot)$ is the density function of $t_{\nu}$ and $\pi_j$ is the initial combining weight of the $j$-th
forecaster. So, $q^n$ is the estimator of $f^n$.

Then, for any $1\leq j'\leq J$,
\begin{align*}
\log(f^n/q^n) & \leq \log\frac{\prod_{i=i_0+1}^{i_0+n}\frac{1}{s_i}h\bigl(\frac{y_i-m_i}{s_i}\bigl)}{
\pi_{j'}\prod_{i=i_0+1}^{i_0+n}\frac{1}{\hat{s}_{i,j'}}h\bigl(\frac{y_i-\hat{y}_{i,j'}}{\hat{s}_{i,j'}}\bigl)}
= \log\frac{1}{\pi_{j'}} +
\sum_{i=i_0+1}^{i_0+n}\log\frac{\frac{1}{s_i}h\bigl(\frac{y_i-m_i}{s_i}\bigl)}{\frac{1}{\hat{s}_{i,j'}}
h\bigl(\frac{y_i-\hat{y}_{i,j'}}{\hat{s}_{i,j'}}\bigl)}
\end{align*}

Conditional on all the information before time point $i$, 
\begin{align*}
E_i \log\frac{\frac{1}{s_i}h\bigl(\frac{Y_i-m_i}{s_i}\bigl)}{\frac{1}{\hat{s}_{i,j'}}
h\bigl(\frac{Y_i - \hat{y}_{i,j'}}{\hat{s}_{i,j'}}\bigl)}
& = \int \frac{1}{s_i}h\bigl(\frac{y_i-m_i}{s_i}\bigl)\log\frac{\frac{1}{s_i}h\bigl(\frac{y_i-m_i}{s_i}\bigl)}{
\frac{1}{\hat{s}_{i,j'}} h\bigl(\frac{y_i-\hat{y}_{i,j'}}{\hat{s}_{i,j'}}\bigl)} dy_i \\
& = \int h(x)
\log\frac{h(x)}{\frac{1}{\hat{s}_{i,j'}/s_i}h\bigl(\frac{x-(\hat{y}_{i,j'}-m_i)/s_i}{\hat{s}_{i,j'}/s_i}
\bigl)}dx
\end{align*}
By the Lemma 1 in A.2,
\begin{equation*}
E_i \log\frac{\frac{1}{s_i}h\bigl(\frac{Y_i-m_i}{s_i}\bigl)}{\frac{1}{\hat{s}_{i,j'}}
h\bigl(\frac{Y_i - \hat{y}_{i,j'}}{\hat{s}_{i,j'}}\bigl)} \leq
\frac{(\hat{y}_{i,j'}-m_i)^2}{2s_i^2} +
B_1\frac{(\hat{s}_{i,j'}-s_i)^2}{s_i^2}
\end{equation*}
where $B_1 = \nu\frac{2+s_0}{2s_0}$. So, 
\begin{equation*}
\frac{1}{n}\sum_{i=i_0+1}^{i_0+n} E D(q_i||\hat{q}_i^{A_t})
\leq 
\inf_{1\leq j\leq J}\left(\frac{\log\frac{1}{w_j^{A_t}}}{n} +
\frac{1}{n}\sum_{i=i_0+1}^{i_0+n}E\frac{(\hat{y}_{i,j}-m_i)^2}{2s_i^2}
+ \frac{B_1}{n}\sum_{i=i_0+1}^{i_0+n} E\frac{(\hat{s}_{i,j}-s_i)^2}{s_i^2}\right)
\end{equation*}

From the Theorem 1 of \citet{Y2004}, there exists a constant $C$ depending on the parameters in Conditions 1 and 2$'$, such that,
\begin{equation*}
E D(q_i||\hat{q}_i^{A_t})\geq \frac{1}{C}E \frac{(m_i-\hat{y}_i^{A_t})^2}{\sigma_i^2}.
\end{equation*}
Therefore,
\begin{equation*}
\frac{1}{n}\sum_{i=i_0+1}^{i_0+n} E \frac{(m_i-\hat{y}_i^{A_t})^2}{\sigma_i^2}
\leq 
C\inf_{1\leq j\leq J}\left(\frac{\log\frac{1}{w_j^{A_t}}}{n} +
\frac{B_2}{n}\sum_{i=i_0+1}^{i_0+n}E\frac{(\hat{y}_{i,j}-m_i)^2}{\sigma_i^2}
+ \frac{B_3}{n}\sum_{i=i_0+1}^{i_0+n} E\frac{(\hat{s}_{i,j}-s_i)^2}{s_i^2}\right),
\end{equation*}
where $B_2$ is a function of $\nu$ and $B_3$ is deducted the same as $B_1$ but under Condition 2$'$ instead of Condition 2.

\subsection*{A.4}

Essential part of the proof of Theorem 2 is provided in this subsection. We only provide the steps of the proof when the random errors are scaled Student's $t$-distributed since proof of other situations are similar.

Let $\hat{s}_{i,j,k}$ be the estimator of $s_i$ from the $j$-th forecaster assuming $\nu_k$ is the true degrees of freedom. If Condition 4 holds, then obviously
\begin{equation*}
q^n \geq
\sum_{k=1}^{K}\sum_{j=1}^Jc_2w_{j,k}^{A_t}/G\prod_{i=i_0+1}^{i_0+n}\frac{1}{\hat{s}_{i,j,k}}h_{\nu_l}\bigg(\frac{y_i-\hat{y}_{i,j}} { \hat{s}_{i,j,k}} \biggl).
\end{equation*}

So, for any $j^{\ast}$ and $k^{\ast}$,
\begin{align*}
\log\frac{f^n}{q^n} & \leq \log\frac{\prod_{i=i_0+1}^{i_0+n}\frac{1}{s_i}h\bigl(\frac{y_i-m_i}{s_i}\bigl)}{
c_2w_{j^{\ast},k^{\ast}}^{A_t}/G\prod_{i=i_0+1}^{i_0+n}\frac{1}{\hat{s}_{i,j^{\ast},k^{\ast}}}h_{\nu_{k^{\ast}}}\bigg(\frac{y_i-
\hat{y }_{i,j^{\ast}}} { \hat{s}_{i,j^{\ast},k^{\ast}}} \biggl)}
= \log\frac{G}{c_2w_{j^{\ast},k^{\ast}}^{A_t}} +
\sum_{i=i_0+1}^{i_0+n}\log\frac{\frac{1}{s_i}h\bigl(\frac{y_i-m_i}{s_i}\bigl)}{\frac{1}{\hat{s}_{i,j^{\ast},k^{\ast}}}
h\bigl(\frac{y_i-\hat{y}_{i,j^{\ast}}}{\hat{s}_{i,j^{\ast},k^{\ast}}}\bigl)}.
\end{align*}
Similarly, by the Lemma 1 in A.2,
\begin{equation*}
E_i \log\frac{\frac{1}{s_i}h\bigl(\frac{Y_i-m_i}{s_i}\bigl)}{\frac{1}{\hat{s}_{i,j^{\ast},k^{\ast}}}
h\bigl(\frac{Y_i-\hat{y}_{i,j^{\ast}}}{\hat{s}_{i,j^{\ast},k^{\ast}}}\bigl)} \leq
B_1\frac{(\hat{y}_{i,j^{\ast}}-m_i)^2}{\sigma_i^2} + B_2\frac{(\hat{s}_{i,j^{\ast},k^{\ast}}-s_i)^2}{s_i^2} +
B_3\bigl|\frac{\nu_k-\nu}{\nu}\bigl|.
\end{equation*}
The rest of the proof is similar to that of Theorem 1. 

\bibliographystyle{mdpi}
\makeatletter
\renewcommand\@biblabel[1]{#1. }
\makeatother

\begin{table*}
\begin{center}
\caption{Simulation Results on the Linear Regression Models}
\label{Table:1}
\begin{tabular}{c|cc|cc|cc|cc}   \hline
      & \multicolumn{2}{c|}{$t_3$}    & \multicolumn{2}{c|}{$DE$} 
      & \multicolumn{2}{c|}{$t_{10}$} & \multicolumn{2}{c}{$normal$}      \\ \hline
      & $\sigma^2=1$  & $\sigma^2=9$   & $\sigma^2=1$  & $\sigma^2=9$  
      & $\sigma^2=1$  & $\sigma^2=9$   & $\sigma^2=1$  & $\sigma^2=9$    \\ \hline
      
      & \multicolumn{8}{c}{$p_0=3$} \\ \hline
$A2$	&	1.302	&	1.043	&	1.116	&	1.028	&	0.983	&	0.958	&	0.926	&	0.931	\\[-1ex]	
	&	(0.009)	&	(0.003)	&	(0.004)	&	(0.001)	&	(0.003)	&	(0.001)	&	(0.002)	&	(0.001)	\\	\hline
$At$	&	0.943	&     0.980	        &      0.983	&      0.995	&      0.941	&	0.955	&	0.932	&	0.942	\\	[-1ex]
	&	(0.002)	& 	(0.001)	&	(0.001)	&	(0.001)	&	(0.003)	&	(0.001)	&	(0.001)	&	(0.001)	\\	\hline
$Ag$	&	0.944	&     0.967	        &      0.974	&  	0.977	&      0.940	&	0.950	&	0.926	&	0.938	\\	[-1ex]
	&	(0.002)	&	(0.001)	&	(0.001)	&	(0.001)	&	(0.001)	&	(0.001)	&	(0.001)	&	(0.001)	\\	\hline

      & \multicolumn{8}{c}{$p_0=5$} \\ \hline
$A2$	&	1.257	&	1.066	&	1.088	&	1.026	&	0.980	&	0.955	&	0.937	&	0.927	\\[-1ex]	
	&	(0.008)	&	(0.004)	&	(0.003)	&	(0.001)	&	(0.002)	&	(0.001)	&	(0.002)	&	(0.001)	\\	\hline
$At$	&	0.950	&	0.967	&	0.976	&	0.982	&	0.951	&	0.950	&	0.943	&	0.938	\\	[-1ex]
	&	(0.002)	&	(0.001)	&	(0.001)	&	(0.001)	&	(0.001)	&	(0.001)	&	(0.001)	&	(0.001)	\\	\hline
$Ag$	&	0.951	&	0.958	&	0.971	&	0.970	&	0.949	&	0.944	&	0.939	&	0.933	\\	[-1ex]
	&	(0.001)	&	(0.001)	&	(0.001)	&	(0.001)	&	(0.001)	&	(0.001)	&	(0.001)	&	(0.001)	\\	\hline
      
      & \multicolumn{8}{c}{$p_0=10$} \\ \hline      
$A2$	&	1.166	&	1.056	&	1.035	&	0.998	&	0.968	&	0.949	&	0.946	&	0.929	\\[-1ex]	
	&	(0.006)	&	(0.003)	&	(0.002)	&	(0.001)	&	(0.002)	&	(0.001)	&	(0.001)	&	(0.001)	\\	\hline
$At$	&	0.950	&	0.957	&	0.964	&	0.965	&	0.949	&	0.946	&	0.948	&	0.939	\\	[-1ex]
	&	(0.002)	&	(0.001)	&	(0.001)	&	(0.001)	&	(0.001)	&	(0.001)	&	(0.001)	&	(0.001)	\\	\hline
$Ag$	&	0.945	&	0.949	&	0.961	&	0.955	&	0.944	&	0.939	&	0.942	&	0.933	\\	[-1ex]
	&	(0.001)	&	(0.001)	&	(0.001)	&	(0.001)	&	(0.001)	&	(0.001)	&	(0.001)	&	(0.001)	\\	\hline
\end{tabular}   
\end{center}
\end{table*}

\begin{table*}
\begin{center}
\caption{Simulation Results on the $AR$ Models with $p=5$ (not or only mildly heavy tailed)}
\label{Table:2}
\begin{tabular}{c|ccc|ccc|ccc}   \hline
            & \multicolumn{3}{c|}{$normal$}  & \multicolumn{3}{c|}{$t_{10}$} & \multicolumn{3}{c}{$DE$}  \\ \hline
            & $\sigma^2=1$  & $\sigma^2=4$  & $\sigma^2=9$   
            & $\sigma^2=1$  & $\sigma^2=4$  & $\sigma^2=9$  
            & $\sigma^2=1$  & $\sigma^2=4$  & $\sigma^2=9$        \\ \hline
$A2$	&	0.941 	&	0.940 	&	0.940 	&	0.972 	&	0.972 	&	0.971 	&	1.030 	&	1.032 	&	1.033 	\\[-1ex]	
	&	(0.004)	&	(0.004)	&	(0.004)	&	(0.004)	&	(0.003)	&	(0.003)	&	(0.004)	&	(0.003)	&	(0.004)	\\	\hline
$At$	&	0.954 	&	0.953 	&	0.954 	&	0.961 	&	0.962 	&	0.962 	&	0.997 	&	1.001 	&	0.995 	\\	[-1ex]
	&	(0.003)	&	(0.003)	&	(0.003)	&	(0.002)	&	(0.003)	&	(0.003)	&	(0.001)	&	(0.001)	&	(0.001)	\\	\hline
$Ag$	&	0.948 	&	0.947 	&	0.948 	&	0.957 	&	0.959 	&	0.958 	&	0.978 	&	0.983 	&	0.976 	\\[-1ex]	
	&	(0.003)	&	(0.004)	&	(0.004)	&	(0.003)	&	(0.003)	&	(0.003)	&	(0.002)	&	(0.001)	&	(0.002)	\\	\hline
$SA$	&	2.892 	&	2.484 	&	2.408 	&	2.372 	&	2.297 	&	2.070 	&	2.278 	&	2.176 	&	2.483 	\\[-1ex]	
	&	(0.268)	&	(0.166)	&	(0.189)	&	(0.167)	&	(0.174)	&	(0.127)	&	(0.148)	&	(0.151)	&	(0.148)	\\	\hline
$MD$&	1.681 	&	2.025 	&	1.824 	&	1.884 	&	1.874 	&	1.421 	&	1.740 	&	1.602 	&	1.943 	\\	[-1ex]
	&	(0.137)	&	(0.191)	&	(0.187)	&	(0.243)	&	(0.197)	&	(0.076)	&	(0.137)	&	(0.144)	&	(0.168)	\\	\hline
$TM$&	1.805 	&	1.946 	&	1.754 	&	1.838 	&	1.705 	&	1.469 	&	1.723 	&	1.571 	&	1.885 	\\	[-1ex]
	&	(0.121)	&	(0.144)	&	(0.134)	&	(0.156)	&	(0.138)	&	(0.066)	&	(0.109)	&	(0.093)	&	(0.120)	\\	\hline
$BG$&	1.441 	&	1.462 	&	1.389 	&	1.425 	&	1.364 	&	1.321 	&	1.431 	&	1.357 	&	1.500 	\\[-1ex]	
	&	(0.047)	&	(0.051)	&	(0.047)	&	(0.042)	&	(0.040)	&	(0.032)	&	(0.046)	&	(0.035)	&	(0.045)	\\	\hline
$BG_{0.95}$&	1.432 &	1.453 	&	1.381 	&	1.417 	&	1.358 	&	1.315 	&	1.427 	&	1.353 	&	1.495 	\\	[-1ex]
	            &(0.047)&   (0.050)	&	(0.047)	&	(0.042)	&	(0.040)	&	(0.032)	&	(0.045)	&	(0.035)	&	(0.045)	\\	\hline
$BG_{0.9}$&	1.429&	1.449 	&	1.378 	&	1.414 	&	1.355 	&	1.313 	&	1.425 	&	1.352 	&	1.492 	\\	[-1ex]
	&	(0.047)	&	(0.049)	&	(0.047)	&	(0.042)	&	(0.039)	&	(0.032)	&	(0.045)	&	(0.035)	&	(0.045)	\\	\hline
$BG_{0.8}$&1.433 	&	1.452 	&	1.382 	&	1.417 	&	1.357 	&	1.315 	&	1.427 	&	1.353 	&	1.491 	\\	[-1ex]
	&(0.047)            &(0.050)	       &	(0.047)	&	(0.042)	&	(0.040)	&	(0.032)	&	(0.045)	&	(0.035)	&	(0.044)	\\	\hline
$BG_{0.7}$&1.447 	&	1.464 	&	1.394 	&	1.428 	&	1.366 	&	1.322 	&	1.432 	&	1.357 	&	1.495 	\\	[-1ex]
	&	(0.048)	&	(0.051)	&	(0.049)	&	(0.043)	&	(0.040)	&	(0.033)	&	(0.046)	&	(0.036)	&	(0.045)	\\	\hline
$LR$&	7.956 	&	8.355 	&	8.491 	&	8.856 	&	10.210 	&	9.138 	&	11.110 	&	11.240 	&	10.040 	\\[-1ex]	
	&	(0.346)	&	(0.339)	&	(0.342)	&	(0.387)	&	(1.032)	&	(0.363)	&	(0.504)	&	(0.509)	&	(0.513)	\\	\hline
$CLR$&	1.036 	&	1.024 	&	1.036 	&	1.032 	&	1.036 	&	1.042 	&	1.072 	&	1.070 	&	1.045 	\\	[-1ex]
	&	(0.011)	&	(0.013)	&	(0.012)	&	(0.011)	&	(0.010)	&	(0.011)	&	(0.011)	&	(0.011)	&	(0.013)	\\	\hline
\end{tabular}     
\end{center}
\end{table*}

\begin{table*}
\begin{center}
\caption{Simulation Results on the $AR$ Models with $p=5$ (heavy tailed)}
\label{Table:3}
\begin{tabular}{c|ccc|ccc}   \hline
& \multicolumn{3}{c|}{$t_{3}$} & \multicolumn{3}{c}{log-normal} \\ \hline
            & $\sigma^2=1$  & $\sigma^2=4$  & $\sigma^2=9$ 
            & $\sigma=0.25$ & $\sigma=0.5$  & $\sigma=1$         \\ \hline
$A2$	&	1.058 	&	1.056 	&	1.053 	&	0.964 	&	1.024 	&	1.051 	\\[-1ex]	
	&	(0.009)	&	(0.008)	&	(0.008)	&	(0.003)	&	(0.004)	&	(0.010)	\\	\hline
$At$	&	0.955 	&	0.947 	&	0.961 	&	0.951 	&	0.940 	&	0.921 	\\	[-1ex]
	&	(0.006)	&	(0.006)	&	(0.006)	&	(0.003)	&	(0.004)	&	(0.008)	\\	\hline
$Ag$	&	0.950 	&	0.943 	&	0.957 	&	0.950 	&	0.946 	&	0.926 	\\	[-1ex]
	&	(0.006)	&	(0.006)	&	(0.006)	&	(0.003)	&	(0.004)	&	(0.008)	\\	\hline
$SA$	&	2.047 	&	1.889 	&	1.931 	&	2.253 	&	2.143 	&	1.730 	\\	[-1ex]
	&	(0.107)	&	(0.098)	&	(0.139)	&	(0.173)	&	(0.115)	&	(0.087)	\\	\hline
$MD$	&	1.692 	&	1.396 	&	1.657 	&	1.517 	&	1.441 	&	1.370 	\\	[-1ex]
	&	(0.135)	&	(0.066)	&	(0.182)	&	(0.097)	&	(0.085)	&	(0.078)	\\	\hline
$TM$	&	1.625 	&	1.438 	&	1.508 	&	1.559 	&	1.555 	&	1.404 	\\	[-1ex]
	&	(0.091)	&	(0.060)	&	(0.112)	&	(0.086)	&	(0.080)	&	(0.057)	\\	\hline
$BG$	&	1.369 	&	1.307 	&	1.286 	&	1.329 	&	1.374 	&	1.278 	\\	[-1ex]
	&	(0.034)	&	(0.025)	&	(0.033)	&	(0.039)	&	(0.038)	&	(0.025)	\\	\hline
$BG_{0.95}$	&	1.365 	&	1.303 	&	1.282 	&	1.322 	&	1.370 	&	1.275 	\\	[-1ex]
	&	(0.033)	&	(0.025)	&	(0.033)	&	(0.038)	&	(0.038)	&	(0.025)	\\	\hline
$BG_{0.9}$	&	1.360 	&	1.299 	&	1.277 	&	1.319 	&	1.367 	&	1.271 	\\	[-1ex]
	&	(0.033)	&	(0.025)	&	(0.032)	&	(0.037)	&	(0.037)	&	(0.024)	\\	\hline
$BG_{0.8}$	&	1.352 	&	1.290 	&	1.269 	&	1.320 	&	1.366 	&	1.259 	\\	[-1ex]
	&	(0.032)	&	(0.024)	&	(0.030)	&	(0.038)	&	(0.037)	&	(0.023)	\\	\hline
$BG_{0.7}$	&	1.345 	&	1.284 	&	1.263 	&	1.327 	&	1.368 	&	1.248 	\\	[-1ex]
	&	(0.032)	&	(0.023)	&	(0.030)	&	(0.039)	&	(0.037)	&	(0.023)	\\	\hline
$LR$	&	95.280 	&	38.290 	&	46.220 	&	9.316 	&	13.180 	&	174.000 	\\	[-1ex]
	&	(60.670)	&	(7.566)	&	(9.192)	&	(0.375)	&	(0.891)	&	(56.286)	\\	\hline
$CLR$	&	1.014 	&	1.007 	&	1.016 	&	1.046 	&	1.032 	&	0.974 	\\	[-1ex]
	&	(0.010)	&	(0.010)	&	(0.010)	&	(0.011)	&	(0.011)	&	(0.010)	\\	\hline
\end{tabular} \\
{\tt Note}:  For the columns of `log-normal',  $\sigma$'s are the scale parameters. 
\end{center}
\end{table*}

\begin{table*}
\begin{center}
\caption{Results on the 1428 Variables of the M3-Competition Data}
\label{Table:4}
\begin{tabular}{c|ccccccc}   \hline
            &  mean & se    & median &  min   & $Q_1$  & $Q_3$ & max    \\ \hline
$MD$        & 1.050 & 0.010 & 1.022  & 0.002  & 0.910  & 1.143 & 5.341  \\
$TM$        & 0.990 & 0.004 & 1.000  & 0.002  & 0.974  & 1.023 & 2.437  \\ 
$BG$        & 0.784 & 0.010 & 0.838  & 0.001  & 0.596  & 0.973 & 5.227  \\
$BG_{0.95}$ & 0.775 & 0.010 & 0.832  & 0.001  & 0.582  & 0.969 & 7.715  \\ 
$BG_{0.9}$  & 0.768 & 0.012 & 0.825  & 0.001  & 0.564  & 0.966 & 11.45  \\ 
$BG_{0.8}$  & 0.758 & 0.019 & 0.806  & 0.001  & 0.529  & 0.960 & 24.08  \\ 
$BG_{0.7}$  & 0.757 & 0.031 & 0.793  & 0.001  & 0.503  & 0.956 & 43.19  \\ 
$A1$        & 0.708 & 0.016 & 0.649  & 0.001  & 0.307  & 0.994 & 11.50  \\ 
$A2$        & 0.697 & 0.017 & 0.639  & 0.001  & 0.309  & 0.979 & 13.32  \\
$At$        & 0.708 & 0.015 & 0.646  & 0.001  & 0.312  & 1.003 & 8.632  \\ 
$Ag$        & 0.696 & 0.014 & 0.645  & 0.001  & 0.308  & 0.987 & 7.710  \\  \hline
\end{tabular}  
\end{center}
\end{table*}

\clearpage
\begin{table*}
\begin{center}
\caption{Results on the Heavy-tailed Subset}
\label{Table:5}
\begin{tabular}{c|ccccccc}   \hline
      &  mean & se    & median &  min   & $Q_1$  & $Q_3$ & max    \\ \hline
$SA$  &  7.738 &   1.695  &  2.259 &  0.131 &  1.311 &  5.244 &   82.734 \\ 
$MD$ &   8.088 &  2.005  &  1.912 &  0.222 &  1.162 &  4.974 &  120.428 \\ 
$TM $ &  7.607 &  1.664  &  2.299 &  0.129 &  1.267 &  5.175  &  78.481 \\ 
$BG_{0.95}$  &  2.017 &  0.217  &  1.431 &  0.241 &  0.965 &  2.472  &  12.551 \\ 
$BG_{0.9}$  &  1.846 &  0.182  &  1.337 &  0.208 &  0.958 &  2.444  &  10.383 \\ 
$BG_{0.8}$  &  1.656 &  0.150  &  1.340 &  0.179 &  0.851 &  2.074  &   8.577 \\ 
$BG _{0.7}$ &  1.536 &  0.141  &  1.256 &  0.158 &  0.813 &  1.673  &   7.746 \\  \hline
\end{tabular}  
\end{center}
\end{table*}

\end{document}